%% file: gx304.tex
\documentclass{aa}  

\usepackage{graphicx}
\usepackage{subfigure}
\usepackage{float}
\usepackage[caption = false]{subfig}
\usepackage{txfonts}
\usepackage{natbib}
\usepackage{inputenc}
\usepackage{longtable, pdflscape, multirow, booktabs}
\usepackage{balance}
\usepackage[dvips]{color}
\usepackage[normalem]{ulem}
\usepackage{fancybox}

\def \sw {{\it Swift}}

\def \src {\mbox{GX~$304$-$1$~}}

\begin{document}

   \title{Luminosity-dependent spectral and timing properties of the
     accreting pulsar GX 304$-$1 measured with INTEGRAL}
   \titlerunning{INTEGRAL observations of \src}
   \author{C. Malacaria
          \inst{}
          \and
          D. Klochkov
          \and
          A. Santangelo
          \and
          R. Staubert
          }

   \institute{Institut f\"{u}r Astronomie und Astrophysik, Sand 1, 72076 T\"{u}bingen, Germany\\
              \email{malacaria@astro.uni-tuebingen.de}}
         %\and
          %   University of Alexandria, Department of Geography, ...\\
           %  \email{c.ptolemy@hipparch.uheaven.space}
            % \thanks{The university of heaven temporarily does not
             %        accept e-mails}
             %}

   \date{\today}

\abstract
{Be/X-ray binaries show outbursts with peak luminosities up to a few times $10^{37}$ erg/s, during which they can be observed and studied in detail. Most (if not all) Be/X-ray binaries harbour accreting pulsars, whose X-ray spectra in many cases contain cyclotron resonant scattering features related to the magnetic field of the sources. Spectral variations as a function of luminosity and of  the rotational phase of the neutron star are observed in many accreting pulsars.} 
{We explore X-ray spectral and timing properties of the Be/X-ray binary \src during an outburst episode. Specifically, we investigate the behavior of the cyclotron resonant scattering feature, the continuum spectral parameters, the pulse period, and the energy- and luminosity-resolved pulse profiles. We combine the luminosity-resolved spectral and timing analysis to probe the accretion geometry and the beaming patterns of the rotating neutron star.} 
{We analyze the INTEGRAL data from the two JEM-X modules, ISGRI and SPI, covering the January-February 2012 outburst, divided in six observations. We obtain pulse profiles in two energy bands, phase-averaged and phase-resolved spectra for each observation.} 
{We confirm the positive luminosity-dependence of the cyclotron line energy in GX~304-1, and report a dependence of the photon index on luminosity. Using a pulse-phase connection technique, we find a pulse period solution valid for the entire outburst. Our pulse-phase resolved analysis shows, that the centroid energy of the cyclotron line is varying only slightly with pulse phase, while other spectral parameters show more pronounced variations. Our results are consistent with a scenario in which, as the pulsar rotates, we are exploring only a small portion of its beam pattern.} 
{}


   \maketitle
%
%________________________________________________________________

\section{Introduction}
\input{Chapters/1_Introduction.tex}
\label{sec:intro}

\section{Observations and data}
\input{Chapters/2_Data.tex}
\label{sec:observation}

\section{Timing analysis}
\input{Chapters/3_Timing.tex}
\label{sec:timing}

\section{Spectral analysis}
\input{Chapters/4_Spectral.tex}

\label{sec:spectral}

\section{Phase-resolved spectral analysis}
\input{Chapters/5_prs.tex}
\label{sec:analysis:prs}

\section{Discussion}
\input{Chapters/6_Discussion.tex}
\label{sec:discussion}

\section{Summary and conclusions}
\input{Chapters/7_Conclusions.tex}
\label{sec:conclusions}

\begin{acknowledgements}
We gratefully acknowledge the anonymous referee for numerous comments that greatly improved the manuscript.
This research is based on observations with \textit{INTEGRAL}, an ESA project with the instruments
and science data center funded by ESA member states especially  the PI countries: 
Denmark, France, Germany, Italy, Switzerland, Spain), Czech Republic and Poland, 
and with the participation  of Russia and the USA.
This work is supported by the \textsl{Bundesministerium f\"{u}r Wirtschaft und Technologie} 
through the \textsl{Deutsches Zentrum f\"{u}r Luft- und Raumfahrt e.V. (DLR)} under the grant number FKZ 50 OR 1204.
C.M. thanks the \textit{MAXI} team (RIKEN, Japan) for the support during our collaboration.
\end{acknowledgements}

%-------------------------------------------------------------------

\bibliographystyle{aa} % style aa.bst
%\bibliography{gx304malacaria}
\bibliography{gx304}

\begin{appendix}
\section{The OSA 10 data reduction}
\input{Chapters/Appendix.tex}\label{sec:app}
\end{appendix}
\balance

\end{document}

%% file: Chapters/1_Introduction.tex
\object{GX~304-1} is a Be/X-ray binary (BeXRB) system discovered as an X-ray source 
in 1967 during a balloon observation \citep{1968Natur.219.1235L, 1968ApJ...152L..49L}. 
Subsequently, the source was established to be an X-ray pulsar with a pulse period of $\sim272$\,s \citep{1977ApJ...216L..15M}. 
A study of the recurrent outburst activity revealed a $\sim132.5$\,d periodicity, 
likely due to the system's orbital period \citep{1983ApJ...273..709P}. 
The optical counterpart of the binary is a B2~Vne star, whose distance has 
been measured to be $2.4\pm0.5$ kpc \citep{1980MNRAS.190..537P}. Since 1980, the source entered an 
X-ray off-state \citep{1986A&A...163...93P}, showing no detectable emission for 28 years. 
The quiescence was interrupted in June 2008, when INTEGRAL detected hard X-ray emission from the source 
\citep{2008ATel.1613....1M}.
Since then, \src lighted up repeatedly, becoming a periodically outbursting X-ray source.
The period of the outbursts after $2009$ is roughly the same as before $1980$, i.e., ${\sim}132.5\,$d. 
The peak luminosities are $\lesssim$10$^{37}\,$erg/s in the $5-100\,$keV energy band.
The origin of the X-ray emission is believed to be accretion of matter from the circumstellar equatorial disk 
around the optical companion onto a magnetized neutron star. 
The strong magnetic field of the accretor ($\sim$10$^{12}\,$G) channels the captured matter towards 
its magnetic poles where X-ray emission originates in an accretion structure.

A Cyclotron Resonant Scattering feature (CRSF), or \textit{cyclotron line}, 
has been detected in the spectrum of \src with a centroid energy of $\sim52$ keV \citep{2011PASJ...63S.751Y}. 
CRSFs are important features in the spectra of accreting pulsars. 
In a strong magnetic field, electron energies corresponding to their motion perpendicular 
to the magnetic field lines are quantized in Landau levels, causing resonant scattering of impinging photons. 
The first such line ever was detected in data from a balloon observation of Her~X-1
\citep{1978ApJ...219L.105T}. 
Nowadays, they have turned out to be rather common in accreting X-ray pulsars, with $\sim20$ objects 
being confirmed cyclotron line sources, with several objects showing multiple lines 
(up to four harmonics in 4U~0115+63, \citealt{Santangelo+99}).
Reviews are given by e.g. \citet{Coburn_etal02, Staubert_03, Heindl_etal04, Terada_etal07, Wilms_12, CaballeroWilms_12}.
The energy of the fundamental line $E_{\rm cyc}$ is directly proportional 
to the magnetic field strength at the emission site, 
$E_{\rm cyc}\sim11.6\times B_{12}(1+z_{g})$ keV, 
where $B_{\rm 12}$ is the magnetic field in units of $10^{12}$ G, and $z_{g}$ is the gravitational redshift.

More recent observations have shown that the cyclotron line energy in \src 
is positively correlated with the observed luminosity \citep{Klochkov+12}. 
Such a positive correlation was first observed in Her~X-1 by \citet{Staubert+07} 
and is now observed also for \src, Vela~X-1 \citep{Fuerst+14} 
and recently also for A~0535+26 \citep{Sartore+15}.

% Fig. 1
\begin{figure}[t!]
\includegraphics[width=\hsize]{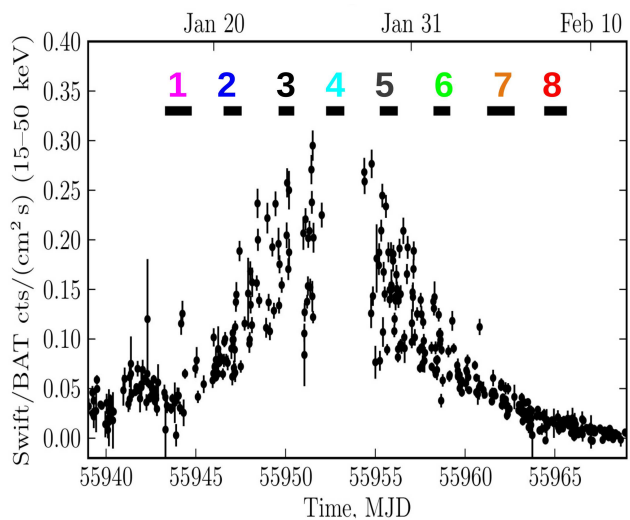}
\caption{The \sw/BAT light curve of \src during the January - February 2012 outburst. 
The horizontal bars mark the INTEGRAL observations. 
The numbers above the bars indicate the last digit 'i' of the respective INTEGRAL revolution number – 113i.}
\label{fig:outburst}
\end{figure}

The opposite correlation, a negative dependence between the cyclotron line energy and the luminosity, 
was actually detected earlier in high luminosity transient sources (4U~0115+63, Cep~X-4, and V~0332+53, 
\citealt{Mihara_98}). 
During the 2004/2005 outburst of V~0332+53, a clear anti-correlation of the line position 
with X-ray flux was observed \citep{Tsygankov_etal06}. 
A similar behavior was observed in outbursts of 4U~0115+63
\citep{Nakajima_etal06, Tsygankov_etal07}.
The presence of the correlation in this source, however, is shown to depend
on the used spectral model \citep{SMueller_etal13, Iyer+15}.

It has also been found that the power law spectral index $\Gamma$ (the absolute value) 
shows the opposite correlation with flux compared to the cyclotron line energy,
both on shorter \citep{2011A&A...532A.126K} and longer \citep{Fuerst+14} time scales.

A model-independent way to probe the correlation of the spectral hardness on flux
is through a hardness-intensity diagram as shown by \citet{2013A&A...551A...1R} for BeXRBs, 
which follow certain patterns in the hardness-intensity 
diagram and undergo state transitions if they reach a luminosity above a certain critical value.
For the physical interpretation of these correlations, that is the association with sub- and super-Eddington
accretion regimes, see \citet{Becker+12} and \citet{2013ApJ...777..115P}.

Recently, \src has passed through a new period of quiescence, 
showing almost no X-ray activity between $2013$ and $2015$.
The data taken during the new active episode allowed \citet{Sugizaki+15} 
to derive orbital elements for this binary system.
Only recently has the source resumed its X-ray activity.

Here we present the results of the timing and spectral studies of \src which was observed by INTEGRAL 
during an outburst in January--February 2012.
Our work is primarily focused on the timing analysis, 
which allowed to phase-connect the observations throughout the entire outburst,
and on the pulse phase-resolved spectroscopy,
with the goal to probe the emission geometry of the rotating NS at different viewing angles.
The X-ray continuum of the source shows variability with pulse phase, 
with some features present only at particular pulse-phase intervals.
Changes of the spectral parameters with pulse phase are quite common among accreting X-ray pulsars 
and are generally attributed to a change in the viewing angle of the emitting region 
(see, e.g., \citealt{Klochkov+08} and \citealt{Vasco+13}).
Located relatively nearby and having a prominent cyclotron line feature,
\src is well suited for such studies.
We confirm the results of \citet{Klochkov+12} using the latest version of the INTEGRAL analysis software and calibration. 
We investigate the continuum variations (photon index and hardness) with luminosity. 
We obtain a pulse period solution that is valid throughout the observed outburst. 
The measured pulse periods are used to study pulse profiles in two energy bands and 
phase-resolved broad band ($5-100\,$keV) spectra.

%% file: Chapters/2_Data.tex
The data used in this work are the same as those used by \citet{Klochkov+12}.
The observation log is given in Table~\ref{table:log}.
The analyzed outburst started on the 8th of January, 2012 \citep{2012ATel.3856....1Y}. 
The International Gamma-ray Astrophysics Laboratory (INTEGRAL, \citealt{2003A&A...411L...1W}) 
observed the event starting around MJD~$55943.5$, when the source flux in the $20-80$ keV energy 
band was $\sim250\,$mCrab, until MJD~$\sim55965.5$, when the flux dropped to $\sim100\,$mCrab.
The peak of the outburst is reached aroun MJD~$55953$, with a flux exceeding $1\,$Crab.
INTEGRAL performed a total of eight observations, one per each satellite orbit 
(i.e., one every about three days), with a typical exposure of a few tens of kiloseconds per each observation.
The three high-energy instruments onboard INTEGRAL allow observations in broad enery ranges: 
the two Joint European X-Ray Monitor units (JEM-X, \citealt{2003A&A...411L.231L}) are sensitive in 
the range $3-35$\,keV; the Imager on Board the INTEGRAL Satellite (IBIS, \citealt{2003A&A...411L.131U}) 
is sensitive from $\sim20$\,keV to a few MeV; and the Spectrometer onboard INTEGRAL (SPI, 
\citealt{2003A&A...411L..63V}) is sensitive in roughly the same energy range as IBIS.
The duration of INTEGRAL observations are shown in Fig.~\ref{fig:outburst}, compared to the \sw/BAT 
light curve of the outburst\footnote{http://swift.gsfc.nasa.gov/results/transients/weak/GX304-1/}. 
The actual observations are shortened due to solar 
activity, leading to a reduction of the exposure times.

\input{Tables/observation_log.tex}
% Fig. 2
\begin{figure}[!ht]
\includegraphics[width=\hsize]{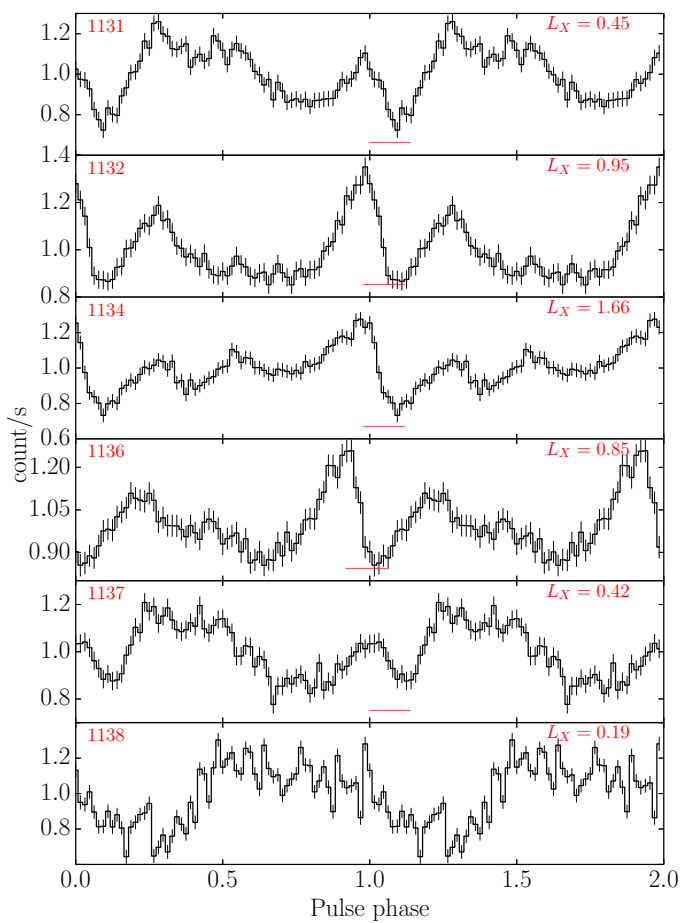}
\caption{The IBIS/ISGRI ($18-80\,$keV) normalized pulse profiles obtained with the timing solution 
	resulting from our phase-connection analysis. 
	Each panel represents, from top to bottom, the successive INTEGRAL 
	observations (from Rev\# $1131$ to $1138$, shown in the top-left corner). 
	In the top-right corner the luminosity of the source ($3-80\,$keV) in units of $10^{37}\,$erg/s is reported.
	The thin straight line around phase $1.0$ at the bottom of each panel 
	(except Rev\# $1138$) approximately marks the phase range where the sharp features 
	used to phase-connect the observations are found.}
\label{fig:isgripp}
\end{figure}
% Fig. 3
\begin{figure}[!ht]
\includegraphics[width=\hsize]{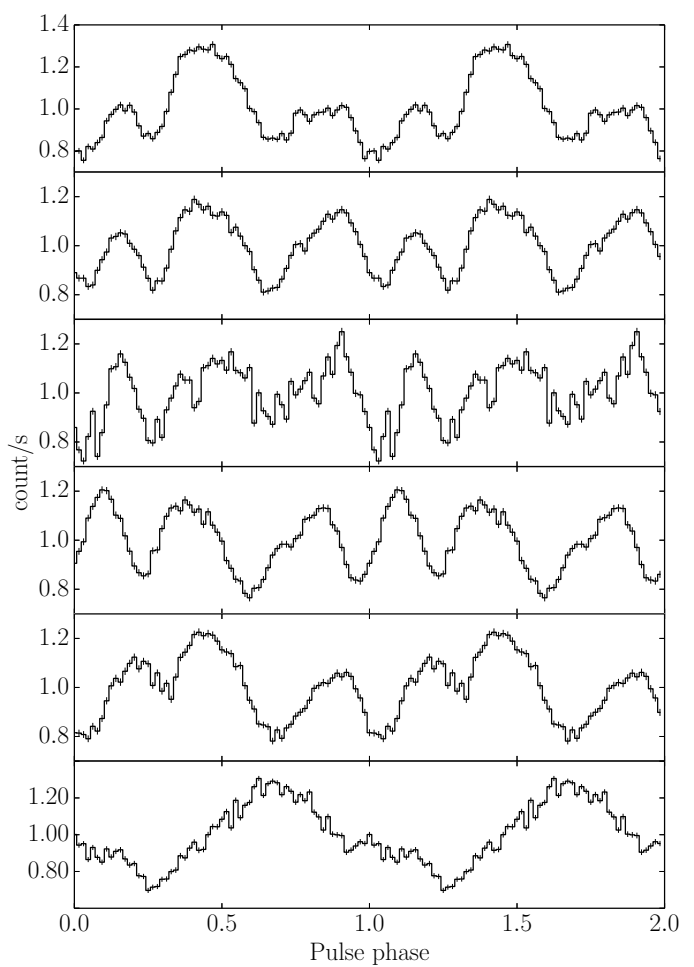}
\caption{JEM-X1 normalized pulse profiles in the $3-20\,$keV energy range.
Panels follow the same Rev\# and luminosity stage as in in Fig.~\ref{fig:isgripp}.}
\label{fig:jmxpp}
\end{figure}

For our analysis, we decided to use only data from those observations where all of the three 
instruments were active, to ensure a broad band coverage and a homogeneous analysis of all observations. 
Therefore, the revolutions 1133 and 1135 are not taken into account.
Of the IBIS instrument, we only used the data from the Integral Soft Gamma-Ray Imager detector 
(ISGRI, \citealt{2003A&A...411L.141L}), which is sensitive in the $20-300\,$keV energy range.
However, according to the recommendation of the instrument team, for the spectral analysis we 
restricted the duty energy band of each instrument to a ``safer'' range which is believed to be free 
of strong systematics: $5-30\,$keV for the two JEM-X modules, $22-100\,$keV for ISGRI and 
$25-100\,$keV for SPI. 
The upper limit of $100\,$keV for ISGRI and SPI data is due to the reduced photon 
flux of \src above this energy. 
Data reduction was performed with the version 10 of the Offline Science Analysis (OSA) 
software\footnote{http://www.isdc.unige.ch/integral/analysis}. 
Following the OSA user manual, we added a systematic error to the final count rates 
at a level of $1\%$ for ISGRI, $3\%$ for JEM-X, and $0.5\%$ for SPI. 
In Section \ref{sec:spectral} also the previous version of the software (OSA 9)
is used to reduce the same data in order to compare the results and to explore the changes 
in the calibration, especially in the ISGRI energy scale, between the two versions.
Details of the comparison between OSA~9 and OSA~10 are given in Appendix~\ref{sec:appendix}.
\input{Tables/timing_values_RSt.tex}\label{table:timing}

%% file: Tables/observation_log.tex
\begin{table}
\caption{INTEGRAL observations of the \src outburst.}\label{table:log}
{\small
\begin{center}
\begin{tabular}{cccccc}
\hline\hline
Rev\# & Obs. ID & Mid. MJD & \multicolumn{3}{c}{Exposure [ks]} \\[0.5ex]
      &         &          &           JEM-X & ISGRI & SPI \\[0.5ex]
\hline
1131 & 09400230006 & 55944.0 & 64.6 & 42.7 & 68.6 \\
1132 & 09400230007 & 55947.0 & 42.4 & 31.9 & 36.6 \\
1133 & 09400230008 & 55950.0 &   –  &  –   & 10.7 \\
1134 & 09400230009 & 55952.8 & 7.3  & 25.4 & 37.8 \\
1135 & 09400230010 & 55955.7 &   –  & 6.7  & 25.1 \\
1136 & 09400230011 & 55958.7 & 36.9 & 28.1 & 32.9 \\
1137 & 09400230012 & 55962.0 & 78.1 & 59.7 & 78.4 \\
1138 & 09400230013 & 55965.0 & 60.7 & 45.2 & 52.3 \\
\hline
\end{tabular}
\end{center}
}
\end{table}

%% file: Tables/timing_values_RSt.tex
\begin{table*}[ht!]
\caption{Global timing solution obtained from phase-connection of IBIS/ISGRI ($18-80\,$keV) pulse-profiles and spectral parameters throughout the outburst obtained with OSA 10 plus additional gain correction (see text). Fluxes and luminosities are calculated in the range $3-80\,$keV.}
\label{table:periods}
\vspace{-4mm}
{\scriptsize
\begin{center}
\begin{tabular}{l c c c c c c c}
\toprule[.25pt]
Rev. ID & Reference values & 1131 & 1132 & 1134 & 1136 & 1137 & 1138 \\[0.5ex]
\toprule[.25pt]
$P$ [s]$^{1}$ & 274.9817 & 275.1313 & 275.0772 & 274.9763 & 274.8714 & 274.8134 & 274.7597 \\[0.5ex]

$T$ [MJD] & 55952.4592 & 55943.98022 & 55947.04829 & 55952.76433 & 55958.70659 & 55961.99595 & 55965.03900 \\[0.5ex]

$Flux/(10^{-8}\,erg\,s^{-1}\,cm^{-2})$ & -- & $0.659_{-0.002}^{+0.002}$ & $1.379_{-0.004}^{+0.004}$ & $2.413_{-0.007}^{+0.006}$ & $1.231_{-0.004}^{+0.004}$ & $0.609_{-0.002}^{+0.002}$ & $0.288_{-0.002}^{+0.002}$\\[0.5ex]

%$Flux/10^{-8}\,erg\,s^{-1}\,cm^{-2}$ & -- & $0.356_{-0.002}^{+0.002}$ & $0.776_{-0.004}^{+0.004}$ & $1.492_{-0.007}^{+0.006}$ & $0.688_{-0.004}^{+0.004}$ & $0.321_{-0.002}^{+0.002}$ & $0.139_{-0.002}^{+0.002}$\\[0.5ex]

$L_{X}/(10^{37}\,erg\,s^{-1})$ & -- & 0.45 & 0.95 & 1.66 & 0.85 & 0.42 & 0.19 \\[0.5ex] % [1ex] adds

%$L_{X}$ & -- & 0.24 & 0.55 & 1.04 & 0.49 & 0.23 & 0.10 \\[0.5ex] % [1ex] adds vertical space

$\Gamma$ & -- & $1.18_{-0.05}^{+0.07}$ & $0.89_{-0.03}^{+0.03}$ & $0.93_{-0.09}^{+0.09}$ & $1.07_{-0.07}^{+0.07}$ & $1.27_{-0.07}^{+0.08}$ & $1.56_{-0.03}^{+0.06}$\\[0.5ex]

$E_{\rm fold} [keV]$ & -- & $21.2_{-0.9}^{+1.2}$ & $16.8_{-0.5}^{+0.6}$ & $16.1_{-0.7}^{+0.7}$ & $17.7_{-0.7}^{+0.8}$ & $21.3_{-1.1}^{+1.2}$ & $30_{-2}^{+2}$\\[0.5ex]

$E_{\rm cyc} [keV]$ & -- & $55.0_{-0.7}^{+0.7}$ & $56.6_{-0.8}^{+0.8}$ & $59.3_{-0.9}^{+0.9}$ & $54.2_{-0.5}^{+0.6}$ & $54.4_{-0.7}^{+0.8}$ & $50.6_{-1.1}^{+1.1}$\\[0.5ex]

$\sigma_{\rm cyc} [keV]$ & -- & $6.6_{-0.5}^{+1.1}$ & $8.5_{-0.4}^{+0.4}$ & $10.8_{-0.8}^{+0.8}$ & $6.5_{-0.4}^{+0.5}$ & $6.5_{-0.5}^{+0.6}$ & $5.3_{-0.7}^{+0.8}$\\[0.5ex]

$\eta_{\rm cyc}$ & -- & $11.9_{-1.4}^{+1.5}$ & $15.7_{-1.8}^{+1.9}$ & $17_{-3}^{+3}$ & $10.5_{-1.1}^{+1.2}$ & $10.4_{-1.3}^{+1.2}$ & $8.1_{-1.5}^{+1.3}$\\[0.5ex]

$\chi^2_{\rm red}$ & -- & 0.9 & 0.8 & 1.3 & 0.9 & 1.0 & 1.0 \\

\bottomrule[.25pt]
\end{tabular}
\\[0.5ex]
\end{center}
}
{\scriptsize$^{1}$ The reference values are the pulse period $P$ valid at the time $T$ in MJD. 
The uncertainties of the pulse periods given is of the order of 0.10 ms.}
%and the epoch-zero time used as reference to calculate the period at the different $T_{0}$ times of different observations. 
% The luminosity is calculated in the $4-80$ keV energy band for a distance of 2.4 kpc and is reported in units of $10^{37}$ erg/s.
\end{table*}

%% file: Chapters/3_Timing.tex
% Fig. 4
\begin{figure}[pt!]
\includegraphics[width=\hsize]{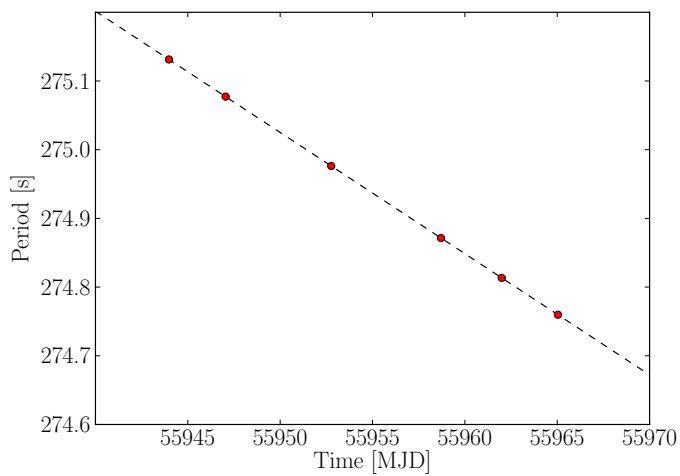}
\caption{Pulse period as a function of time obtained by phase-connection. 
The dashed line represents a linear fit to the data points yielding a pulse period
derivative of $\dot P$ = (-$2.04\pm0.01$)~10$^{-7}$. 
The period error bars are smaller than the symbols.}
\label{fig:phaseconn}
\end{figure}

For the timing analysis, we extracted light curves with a time resolution of $3\,$s from JEM-X1 
(in the range $3-20\,$keV) and IBIS/ISGRI (in the range $18-80\,$keV). 
All time stamps were corrected such that they refer to arrival times at the solar system barycenter. 
For each INTEGRAL observation the pulse period was determined by epoch-folding \citep{1987A&A...180..275L}.
Using these pulse periods, we generated ISGRI pulse profiles in three energy ranges: 
$20-40\,$keV, $40-60\,$keV, and $18-80\,$keV.
The shapes of the profiles do not vary much within these three energy bands. 
Thus, to maximize the photon statistics, we chose the $18-80\,$keV range for further analysis.
At the time of preparation, correction for the orbital motion in the binary was not possible 
since the orbital parameters of the binary system were unknown.
An orbital solution for \src has been furnished in the meanwhile by \citet{Sugizaki+15}.
In our analysis, however, phase connection has been succesfully achieved 
without taking the binary motion into account.
In order to refine the pulse periods necessary to obtain the pulse profiles,
we applied the phase-connection techniques \citep{1981ApJ...247.1003D} to the $18-80\,$keV 
pulse profiles, as described in the following.

For each INTEGRAL observation, a pulse profile was constructed and used to determine an absolute 
reference time (in MJD), the \emph{arrival time}, of the first pulse in the interval by making use 
of well defined 'sharp features' in the pulse profiles, such as - see Fig.~\ref{fig:isgripp} - 
the peak near pulse phase zero and the minimum around pulse phase $0.1$ or the time at the flux level 
centered between the maximum and the minimum.

If only the first derivative $\dot P$ of the pulse period $P$ is taken into account, 
the expected arrival time $t_n$ of the \textit{n}-th pulse is 
\begin{equation}
t_n = t_0 + nP_0 + \frac{1}{2}n^2P\dot{P}.
\label{eq:tn}
\end{equation}
Here, $P_0$ is the pulse period at the reference time $t_0$, \textit{n} is the pulse sequence number, 
and $\dot{P}$ is the first period derivative. 
The pulse period at the time $t$ can be determined as
\begin{equation}\label{cutoffpl}
P(t) = P_0 + (t-t_0) \dot{P}.
\end{equation}

A consistent determination of pulse arrival times is formally only possible if the pulse shape does 
not change with time.

In our observations, however, the pulse shape varies substantially between the intervals. 
Still, the main features of the pulse profile can clearly be recognized in all observations.
The uncertainty of the corresponding pulse arrival times is given by 
the available statistics in the profile for the given shape and sharpness of the feature, 
and in our case corresponds to about $0.01$ in phase.
Using the determined pulse arrival times, the phase connection between the observations can be established.
The pulse period solution ($t_0, P_0, \dot P$) is found from the fit of the measured pulse 
arrival times by formula (\ref{eq:tn}). 
The uncertainties of $P_0$ and $\dot P$ are calculated 
from the $\chi^2$ contours at $1\sigma$ c.l.
This method allows the pulse period evolution to be measured 
with substantially higher precision compared to epoch folding.

% Fig. 5
\begin{figure}[!pt]
    \includegraphics[width=\hsize]{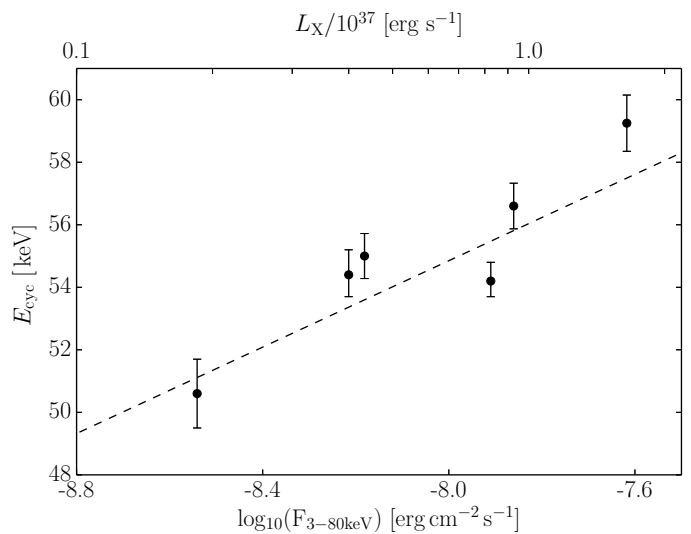}
    \caption{Cyclotron line centroid energy $E_{\rm cyc}$ as a function of the logarithm of flux in the $3-80\,$keV range. 
    The dotted line is the result of a linear fit to the $E_{\rm cyc}-log_{10}(Flux)$ data.
    The error bars indicate $1\sigma$-uncertainties (the flux uncertainties are smaller than the symbol size). 
    The top x-axis shows the corresponding isotropic source luminosity assuming a distance of $2.4\,$kpc.}
    \label{fig:OSA10}
\end{figure}

%Fig. 6
\begin{figure}[!t]
\includegraphics[width=\hsize]{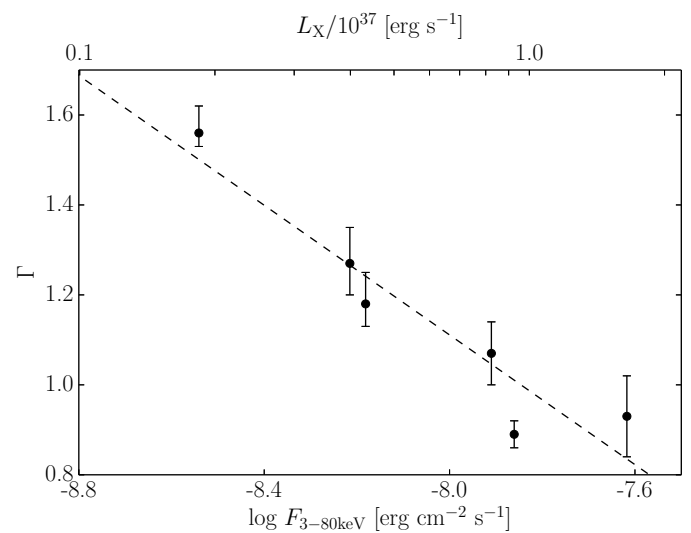}
\caption{Variation of the photon index $\Gamma$ versus the observed flux. 
The dashed line represents the linear fit to the data points. 
The luminosity scale on the top x-axis is as in Fig.~\ref{fig:OSA10}.}
\label{fig:gamma}
\end{figure}

The obtained pulse period solution is presented in Table~\ref{table:periods} and visualized in 
Fig.~\ref{fig:phaseconn}. A decrease of the observed pulse period through the outburst is seen. \\
The initial pulse period is $P_0$ = $274.9817\pm0.0001$ at MJD $55952.4592$ and the period 
derivative is $\dot P$ = (-$2.04\pm0.01$)${\times}10^{-7}$ (constant over the time of observation).
This solution results to be inconsistent with that found by \citet{Sugizaki+15}.
However, such a difference is expected due their different approach, 
which also takes into account the orbital doppler effects.
With the obtained final timing solution we folded the ISGRI light curves to produce a refined set of pulse 
profiles shown in Fig.~\ref{fig:isgripp}. We note that there is still a slight phase shift between the pulse 
profiles, up to $\Delta\phi\sim0.1$, as can be determined, e.g., by the phase of the sharp minimum 
after the largest peak. This is probably due to variations in the pulse profile shape, which is most 
evident in the last observation (see Figs.~\ref{fig:isgripp} and \ref{fig:jmxpp}).
The same timing solution has been used to fold JEM-X1 light curves between $3-20\,$keV and to 
generate the JEM-X pulse profiles for each of the six observations as shown in Fig.~\ref{fig:jmxpp}.

%% file: Chapters/4_Spectral.tex
\label{sec:spectral}

% Fig. 7
\begin{figure}[!t]
\includegraphics[width=\hsize, height=0.74\hsize]{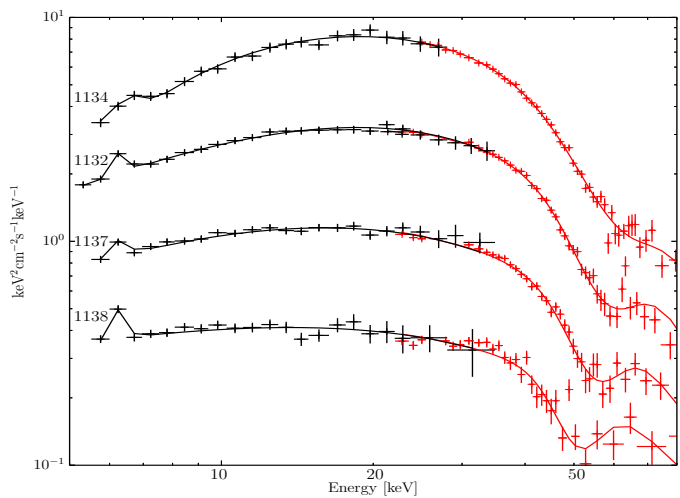}
\caption{Phase-averaged spectra of four observations (labelled above each spectrum), 
from top to bottom, with decreasing luminosity.
JEM-X1 (black) and ISGRI (red) data and folded model are shown.
Spectra have been scaled for better visualization.
Both the $E_{\rm cyc}$--flux and the $\Gamma$--flux correlations are clearly visible.
}
\label{fig:spectra}
\end{figure}

For the following spectral analysis, we used the version $12.7.1$ of the XSPEC software \citep{Arnaud96}.
Following \citet{Klochkov+12}, we fitted the spectrum of \src during the observed outburst 
with a high energy exponential rolloff model (the \texttt{cutoffpl} component in XSPEC):

\begin{equation}\label{cutoffpl}
F(E)\propto E^{-\Gamma}\exp[E/E_{fold}]
\end{equation}

(where E is the photon energy, $\Gamma$ is the photon index and 
$E_{\rm fold}$ is the e-folding energy of exponential rolloff in keV) 
and a multiplicative Gaussian absorption line (the \texttt{gabs} component in XSPEC) 
to account for the CRSF:

\begin{equation}\label{gabs}
G(E)=\text{exp}\left\{-\frac{\eta}{\sqrt{2\pi}\sigma}\text{exp}\left(-\frac{(E-E_{cyc})^2}{2\sigma^2}\right)\right\}
\end{equation}

where $E_{\rm cyc}$ is the cyclotron line centroid energy, $\eta$ and $\sigma$ are the line depth 
(while the optical depth at line center is $\eta/\sqrt{2\pi}/\sigma$), and the width of the line, respectively. 
The $6.4\,$keV Fe $K\alpha$ emission line is modeled with an additive Gaussian, 
even if it does not significantly improve the fit. 
The photo-electric absorption at low energies is modeled with the \texttt{wabs} component in XSPEC.
Using data from both JEM-X modules, ISGRI and SPI, we extracted pulse-phase averaged spectra 
for each observation and explored the behavior of the best-fit parameters over the outburst. 
The results are reported in Table~\ref{table:periods}, and confirm 
the positive correlation of $E_{\rm cyc}$ with flux in GX~304-1. 
The obtained values of the cyclotron line energy show, however, a systematic offset with respect to those 
reported in \citet{Klochkov+12}, which, in turn, is reflected in a higher slope of the fitted line, 
i.e., $6.51\pm1.21\,$keV/log$_{10}$(erg/cm$^2$/s).
The offset results from the new ISGRI calibration implemented in OSA~10.
For technical details about this offset we refer the reader to Appendix~\ref{sec:app}.
The variation of $E_{\rm cyc}$ with flux is shown in Fig.~\ref{fig:OSA10}.

%Fig. 8
\begin{figure}[!t]
\includegraphics[width=\hsize, height=0.74\hsize]{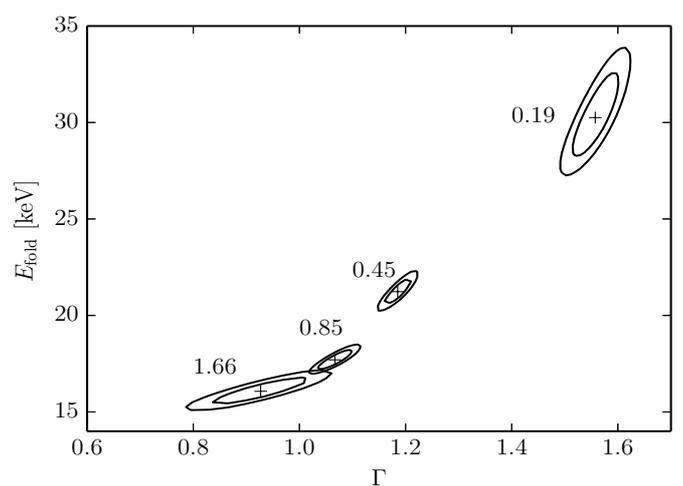}
\caption{$\chi^2$-contour plots of the folding energy $E_{\rm fold}$ and 
the photon index $\Gamma$ for four INTEGRAL observations.
The contours correspond to $\chi^2+1.0$ (the projections of this contour to the parameter axes 
corresponds to the 68\%-uncertainty for one parameter of interest), 
and $\chi^2+2.3$ (68\%-uncertainty for two parameters of interest).
The respective luminosity ($3-80\,$keV) in units of $10^{37}\,$erg\,s$^{-1}$ is indicated.}
\label{fig:contours}
\end{figure}

As mentioned in the Introduction, in addition to the $E_{\rm cyc}$--flux correlation, 
some X-ray pulsars exhibit a correlation between the spectral continuum hardness and flux. 
Using the power law index $\Gamma$ (which describes the spectral hardness below the folding energy), 
an anti-correlation with flux is observed, i.e., the absolute value of the index decreases 
(the spectra become harder) with increasing flux.
This is also true for \src for which we find a negative correlation between the photon index and 
the flux as shown in Fig.~\ref{fig:gamma}.
A selection of spectra is also shown in Fig.~\ref{fig:spectra}, where the spectral variation with luminosity is clear.

For both correlations, $E_{\rm cyc}/\log_{10}({\rm flux})$ and $\Gamma/\log_{10}({\rm flux})$, 
we have determined the Pearson's correlation coefficients $\rho$ and the corresponding one-sided 
probabilities $p$ of obtaining the correlations by chance: $\rho_{\rm cyc}=0.92$, $p_{cyc}\sim0.01$, 
and $\rho_{\rm \Gamma}=-0.95$, $p_{\rm \Gamma}\sim0.004$, respectively.

The folding energy $E_{\rm fold}$ also shows a negative correlation with the observed flux, 
similar to that of the observed photon index.
To verify whether the anti-correlation between the folding energy and the photon index is 
artificial (model-driven), we produced contour plots of the two parameters.
The correlation between the folding energy and the photon index can be seen in Fig.~\ref{fig:contours}.
Altough the contours indicate some intrinsic coupling between $E_{\rm fold}$ and $\Gamma$,
the confidence regions corresponding to different observations are well detached.
Therefore, there must be a physical correlation between the two parameters.
In this plot, $\chi^2$-contours for four observations are also indicated at different luminosities,
showing an anti-correlation between the folding energy and the observed luminosity.

%% file: Chapters/5_prs.tex
\label{sec:prs}

As the neutron star rotates, at each time (or phase) we observe the emitting regions 
of the neutron star at a certain viewing angle. 
To explore the corresponding spectral variability with pulse phase, 
we have extracted the X-ray spectra of \src in five equally spaced pulse-phase intervals
(a finer binning would lead to larger uncertainties).
For this, we have filtered the INTEGRAL data with the Good Time Intervals (GTIs) corresponding to our phase bins.

\begin{figure}[!t]
\includegraphics[angle=-90, width=\hsize]{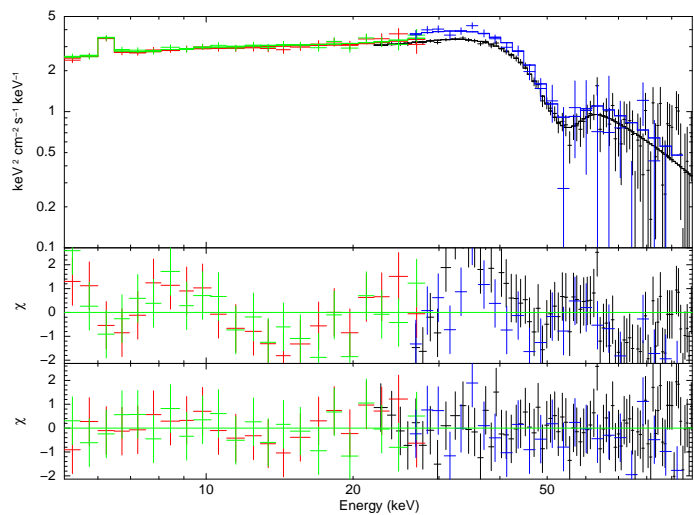}
\caption{\textit{Top:} The spectrum of the phase bin 0.0--0.2 from revolution 1132 fitted with a 
power law/rolloff model including the cyclotron line and an additional Gaussian component to  account 
for a bump around $35\,$keV (see text). 
Both JEM-X (red and green), ISGRI (black) and SPI (blue) data are used.
\textit{Middle:} The residuals for a fit with the model without the additional Gaussian component. 
Large residuals appear around the cut-off energy. 
\textit{Bottom:}The residuals after the inclusion of the Gaussian emission component to model the bump.}  
\label{fig:bump}
\end{figure}

To obtain the phase-resolved spectra, we used the data from both JEM-X modules, ISGRI and SPI.
To fit the phase-resolved spectra, we used the spectral model described in Section \ref{sec:spectral}.
The model provides an acceptable fit (the reduced chi-squared values $\chi^2_{\rm red}$ are between 
0.8 and 1.2) to the spectra from all but one phase bins (see Table \ref{table:prs}). 
The spectra of the 0.0--0.2 phase bins of all revolutions, except the one from Rev\# 1138, however, 
show large residuals with $\chi^2_{\rm red}\gtrsim 1.5$.
Strong positive residuals appear around $35\,$keV.
\input{Tables/prs_results3.tex}\label{table:prs_results}
\input{Tables/bump_parameters2.tex}\label{tab:bump}
These spectra can be modeled with an additional Gaussian emission component (see Fig.~\ref{fig:bump}).
However, to enable a meaningful comparison of the spectral parameters in all phase bins, in Table \ref{table:prs} 
we report the results of the phase-resolved analysis obtained using the same model as in Section \ref{sec:spectral} 
(that is, without the additional Gaussian emission component to model the residuals around $35\,$keV).
A comparison of the models with and without the inclusion of the additional Gaussian line is shown in Table \ref{table:bump}.
Alternatively, instead of the Gaussian emission line to model the bump,
a Gaussian absorption line at lower energies can be used to fit the residuals around $35\,$keV,
resulting in an equally good fit.
In this case, the line centroid energy is about $20\,$keV.
As it can be seen in Table~\ref{table:prs}, the shape of the continuum changes over the pulse phase.
At the same time, the cyclotron line energy $E_{\rm cyc}$ shows only slight variation with pulse phase.
However, the results in Table~\ref{table:prs} show large uncertainties and the variation of many parameters 
with the pulse phase can not be firmly established.
Indeed, even if the nominal value of a given parameter shows some variation, 
it is generally consistent with a constant value.
For a deeper investigation of the spectrum variability with pulse phase, 
we tried to improve the quality of the phase-resolved spectra by stacking together the 
observations at different fluxes.
We therefore have extracted X-ray spectra in $12$ pulse-phase intervals.
The number of phase bins has been chosen to ensure a good statistics in each interval,
while the width of the bin has been kept equal to $0.075$.
In two cases, larger bin sizes are used to obtain a meaningful fit.
Using this approach we loose information on the luminosity dependence 
of the spectra, and possibly introduce some systematic effects 
due to the stacking of observations with different spectral shapes.
However, it results in an improved statistics, and allows us to study 
the flux-averaged dependence of the spectrum on pulse phase.

%Fig. 9

In this analysis we excluded the data of Rev\#~$1138$, which suffers of much lower photon statistics.
The used model for spectral fitting is the same described in Section~\ref{sec:spectral}, which provides 
a good fit ($\chi^2_{\rm red}$ between $0.9$ and $1.2$) for all phase bins except three: 0--0.075, 0.075--0.225 and 0.675--0.75.
Similarly to the 0.0--0.2 phase-bins spectra of the individual observations, 
these phase bins require an additional component to get an acceptable $\chi^2$, 
which can be modeled by a Gaussian emission line around $35\,$keV.
Also here, the residuals in these three phase bins can alternatively be modeled by a Gaussian
absorption line at lower energies, i.e., around $17\,$keV.
Due to the presence of a very similar feature in the phase-resolved spectra of individual observations,
we consider it to be unlikely that such a feature is a result of systematic effects.
However, to ensure a meaningful comparison of the results, we used the same model for all phase bins, 
i.e., the model without the additional Gaussian component.
In order to compare the phase-resolved results with the pulse profile we also produced average pulse profiles 
in the ISGRI and JEM-X energy bands ($18-80$ and $3-20\,$keV, respectively), excluding the data from Rev\#~$1138$.
In Fig.~\ref{fig:totprs} we show the best-fit results of the stacked phase-resolved spectroscopy,
together with the ISGRI and JEM-X average pulse profiles.
The $E_{\rm cyc}$ does not show significant variation with pulse phase 
(except for the phase bin 0.525--0.6), while the photon index $\Gamma$ and the $E_{fold}$ 
confirm a variation which, however, is less than a factor of $\sim2$.

%% file: Tables/prs_results3.tex
\begin{table*}
\caption{Best-fit parameters of the pulse-phase resolved spectra. Fluxes are calculated in the range $3-80\,$keV.}
\label{table:prs}
{\scriptsize
\begin{center}
\begin{tabular}{lccccc|ccccc}
\toprule[.25pt]
\multicolumn{0}{l}{Rev\#} & & & \textbf{1131} & \multicolumn{4}{c}{} & \textbf{1132} \\[0.5ex]
\toprule[.25pt]
Phase-bin & $0.0-0.2*$ & $0.2-0.4$ & $0.4-0.6$ & $0.6-0.8$ & $0.8-1.0$ & $0.0-0.2*$ & $0.2-0.4$ & $0.4-0.6$ & $0.6-0.8$ & $0.8-1.0$\\[0.5ex]
\bottomrule[.25pt]\\[-1.85ex]
$\Gamma$ & ${1.70}_{-0.04}^{+0.04}$ & $0.96_{-0.05}^{+0.05}$ & $1.14_{-0.05}^{+0.12}$ & $1.24_{-0.17}^{+0.16}$ & $1.63_{-0.15}^{+0.13}$ & $1.29_{-0.04}^{+0.08}$ & $0.87_{-0.07}^{+0.04}$ & $1.45_{-0.21}^{+0.81}$ & $0.63_{-0.15}^{+0.17}$ & $0.83_{-0.13}^{+0.13}$ \\[0.5ex]

$E_{\rm fold}$ [keV] & $49_{-45}^{+53}$ & $19_{-1}^{+1}$ & $19_{-2}^{+2}$ & $21_{-2}^{+3}$ & $28_{-3}^{+3}$ & $27_{-1}^{+1}$ & $18_{-1}^{+9}$ & $23_{-3}^{+2}$ & $13_{-1}^{+1}$ & $15_{-1}^{+1}$ \\[0.5ex]

$E_{\rm cyc}$ [keV] & $55_{-3}^{+1}$ & $56_{-1}^{+1}$ & $53_{-1}^{+2}$ & $56_{-2}^{+2}$ & $56_{-1}^{+1}$ & $56.4_{-0.8}^{+0.9}$ & $60_{-2}^{+2}$ & $62_{-3}^{+1}$ & $52_{-2}^{+2}$ & $57_{-1}^{+1}$ \\[0.5ex]

$\sigma_{\rm cyc}$ [keV] & $4.3_{-0.6}^{+0.7}$ & $7.1_{-0.9}^{+1.1}$ & $8.4_{-1.2}^{+1.4}$ & $9.9_{-1.4}^{+1.6}$ & $5.8_{-0.9}^{+0.9}$ & $5.7_{-0.5}^{+0.6}$ & $11_{-2}^{+1}$ & $13_{-3}^{+1}$ & $8.3_{-1.6}^{+2.1}$ & $7.9_{-0.9}^{+1.0}$ \\[0.5ex]

$\eta_{\rm cyc}$ & $10.5_{-1.6}^{+1.8}$ & $15_{-3}^{+4}$ & $13_{-3}^{+4}$ & $19_{-5}^{+5}$ & $14_{-3}^{+3}$ & $16_{-1}^{+2}$ & $22_{-6}^{+6}$ & $26_{-10}^{+6}$ & $10_{-3}^{+5}$ & $18_{-3}^{+4}$ \\[0.5ex]

Flux [$10^{-8}$ erg s$^{-1}$ cm$^{-2}$] & $0.616_{-0.014}^{+0.015}$ & $0.704_{-0.013}^{+0.014}$ & $0.776_{-0.015}^{+0.015}$ & $0.601_{-0.013}^{+0.013}$ & $0.621_{-0.013}^{+0.012}$ & $0.137_{-0.022}^{+0.023}$ & $1.419_{-0.020}^{+0.021}$ & $1.394_{-0.022}^{+0.022}$ & $1.262_{-0.019}^{+0.020}$ & $1.492_{-0.019}^{+0.019}$ \\[0.5ex]

${\chi}^{2}_{\rm red}$ & 1.8 & 1.0 & $0.8$ & $0.9$ & $0.8$ & $2.4$ & $1.0$ & $0.9$ & $0.8$ & $0.8$ \\[0.5ex]

\midrule[.25pt]
\multicolumn{0}{l}{Rev\#} & & & \textbf{1134} & \multicolumn{4}{c}{} & \textbf{1136}\\[0.5ex]
\toprule[.15pt]
Phase-bin & $0.0-0.2*$ & $0.2-0.4$ & $0.4-0.6$ & $0.6-0.8$ & $0.8-1.0$ & $0.0-0.2*$ & $0.2-0.4$ & $0.4-0.6$ & $0.6-0.8$ & $0.8-1.0$\\[0.5ex]
\bottomrule[.25pt]\\[-1.85ex]
$\Gamma$ & $0.59_{-0.06}^{+0.06}$ & $0.39_{-0.15}^{+0.12}$ & $0.88_{-0.20}^{+0.13}$ & $0.21_{-0.18}^{+0.21}$ & $0.59_{-0.18}^{+0.15}$ & $1.56_{-0.03}^{+0.06}$ & $0.80_{-0.05}^{+0.05}$ & $1.07_{-0.18}^{+0.18}$ & $0.83_{-0.16}^{+0.15}$ & $1.06_{-0.12}^{+0.12}$ \\[0.5ex]

$E_{\rm fold}$ [keV] & $17.2_{-0.6}^{+0.6}$ & $13.5_{-1.1}^{+0.8}$ & $17_{-1}^{+1}$ & $13_{-1}^{+1}$ & $13_{-1}^{+1}$ & $33_{-1}^{+2}$& $15.3_{-0.8}^{+0.9}$ & $18_{-2}^{+3}$ & $14_{-1}^{+1}$ & $18_{-1}^{+1}$\\[0.5ex]

$E_{\rm cyc}$ [keV] & $58_{-1}^{+1}$ & $57_{-1}^{+1}$ & $62_{-1}^{+1}$ & $61_{-3}^{+2}$ & $58_{-1}^{+1}$ & $55.1_{-1.1}^{+0.8}$& $52_{-1}^{+1}$ & $58_{-3}^{+3}$ & $51_{-1}^{+1}$ & $55.7_{-0.8}^{+0.9}$\\[0.5ex]

$\sigma_{\rm cyc}$ [keV] & $6.8_{-0.7}^{+0.7}$ & $10.5_{-1.5}^{+2.3}$ & $14_{-1}^{+2}$ & $17_{-2}^{+2}$ & $10.0_{-1.2}^{+1.6}$ & $4.8_{-0.7}^{+0.7}$& $7.7_{-1.1}^{+1.3}$ & $10.0_{-2.2}^{+3.4}$ & $7.2_{-0.9}^{+0.9}$ & $6.0_{-0.6}^{+0.7}$\\[0.5ex]

$\eta_{\rm cyc}$ & $11_{-1}^{+2}$ & $14_{-4}^{+7}$ & $29_{-6}^{+11}$ & $48._{-14}^{+18}$ & $21_{-4}^{+6}$ &$8.2_{-1.3}^{+1.5}$ & $11_{-3}^{+3}$ & $18_{-7}^{+10}$ & $13_{-2}^{+3}$ & $16_{-2}^{+2}$\\[0.5ex]

Flux [$10^{-8}$ erg s$^{-1}$ cm$^{-2}$] & $1.903_{-0.032}^{+0.033}$ & $2.181_{-0.162}^{+0.042}$ & $2.345_{-0.038}^{+0.040}$ & $2.240_{-0.038}^{+0.032}$ & $2.620_{-0.049}^{+0.049}$ & $1.209_{-0.022}^{+0.022}$& $1.279_{-0.021}^{+0.019}$ & $1.162_{-0.024}^{+0.022}$ & $1.158_{-0.019}^{+0.020}$ & $1.261_{-0.018}^{+0.019}$\\[0.5ex]

${\chi}^{2}_{red}$ & $1.5$ & $1.1$ & $1.0$ & $1.1$ &$1.2$ & $1.9$ & $0.9$ & $0.7$ & $0.9$ & $0.8$\\[0.5ex]

\midrule[.25pt]
\multicolumn{0}{l}{Rev\#} & & & \textbf{1137} & \multicolumn{4}{c}{} & \textbf{1138}\\[0.5ex]
\toprule[.15pt]
Phase-bin & $0.0-0.2*$ & $0.2-0.4$ & $0.4-0.6$ & $0.6-0.8$ & $0.8-1.0$ & $0.0-0.2$ & $0.2-0.4$ & $0.4-0.6$ & $0.6-0.8$ & $0.8-1.0$\\[0.5ex]
\bottomrule[.25pt]\\[-1.85ex]
$\Gamma$ & $1.56_{-0.04}^{+0.10}$ & $1.13_{-0.05}^{+0.05}$ & $1.18_{-0.09}^{+0.19}$ & $1.51_{-0.14}^{+0.14}$ & $1.54_{-0.15}^{+0.14}$ & $2.07_{-0.29}^{+0.23}$ & $1.82_{-0.08}^{+0.09}$ & $1.67_{-0.07}^{+0.07}$ & $1.47_{-0.12}^{+0.24}$ & $1.47_{-0.08}^{+0.22}$ \\[0.5ex]

$E_{\rm fold}$ [keV] & $38_{-2}^{+5}$ & $20_{-1}^{+1}$ & $18_{-1}^{+4}$ & $24_{-3}^{+3}$ & $22_{-2}^{+3}$ & $39_{-12}^{+21}$& $77_{-18}^{+42}$ & $37_{-4}^{+5}$ & $22_{-3}^{+5}$ & $29_{-4}^{+7}$\\[0.5ex]

$E_{\rm cyc}$ [keV] & $54.1_{-0.7}^{+0.9}$ & $53_{-1}^{+1}$ & $56_{-2}^{+4}$ & $56_{-2}^{+2}$ & $56_{-2}^{+2}$ & $52_{-3}^{+7}$& $53_{-2}^{+2}$ & $50_{-1}^{+1}$ & $45_{-1}^{+1}$ & $53_{-3}^{+3}$\\[0.5ex]

$\sigma_{\rm cyc}$ [keV] & $4.5_{-0.7}^{+0.6}$ & $5.7_{-0.9}^{+1.0}$ & $10.1_{-1.8}^{+3.9}$ & $8.2_{-1.2}^{+1.4}$ & $5.7_{-1.9}^{+1.8}$ & $5.9_{-4.9}^{+7.0}$& $4.7_{-1.7}^{+1.7}$ & $3.4_{-0.8}^{+0.9}$ & $2.8_{-1.6}^{+1.3}$ & $8.4_{-1.8}^{+1.8}$\\[0.5ex]

$\eta_{\rm cyc}$ & $9.6_{-1.4}^{+1.5}$ & $10.5_{-1.9}^{+2.4}$ & $12.7_{-4.3}^{+10.9}$ & $17_{-4}^{+5}$ & $8.4_{-2.8}^{+3.4}$ &$7.7_{-5.2}^{+12.9}$ & $7.5_{-2.9}^{+3.5}$ & $6.7_{-1.9}^{+2.2}$ & $4.2_{-1.5}^{+1.7}$ & $17.2_{-6.2}^{+7.2}$\\[0.5ex]

Flux [$10^{-8}$ erg s$^{-1}$ cm$^{-2}$] & $0.563_{-0.012}^{+0.006}$ & $0.685_{-0.013}^{+0.013}$ & $0.645_{-0.019}^{+0.028}$ & $0.531_{-0.011}^{+0.011}$ & $0.587_{-0.011}^{+0.012}$ & $0.265_{-0.025}^{+0.018}$& $0.209_{-0.010}^{+0.011}$ & $0.266_{-0.016}^{+0.028}$ & $0.354_{-0.013}^{+0.014}$ & $0.294_{-0.012}^{+0.006}$\\[0.5ex]

${\chi}^{2}_{\rm red}$ & $1.9$ & $1.0$ & $1.1$ & $0.9$ &$0.9$ & $0.9$ & $0.9$ & $1.0$ & $0.9$ & $0.9$\\[0.5ex]
\bottomrule[.25pt]
\end{tabular}
\end{center}
\textbf{Notes:} The starred phase-bins mark those spectra that need an additional gaussian emission line around $35\,$keV to get an accettable fit.}
\end{table*}

%% file: Tables/bump_parameters2.tex
\begin{table*}[t!]
\caption{Best-fit parameters of the first phase-bin spectra using a model with and without additional gaussian emission line.}
\label{table:bump}
{\scriptsize
\begin{center}
\begin{tabular}{r|cc|cc|cc|cc|cc}
\toprule[.25pt]
& \multicolumn{2}{c}{\textbf{Rev\# 1131}} & \multicolumn{2}{c}{\textbf{Rev\# 1132}} & \multicolumn{2}{c}{\textbf{Rev\# 1134}} & \multicolumn{2}{c}{\textbf{Rev\# 1136}} & \multicolumn{2}{c}{\textbf{Rev\# 1137}}\\[0.5ex]
\toprule[.25pt]
Parameter & \textit{No Bump} & \textit{Bump} & \textit{No Bump} & \textit{Bump} & \textit{No Bump} & \textit{Bump} & \textit{No Bump} & \textit{Bump} & \textit{No Bump} & \textit{Bump}\\[0.5ex]
\bottomrule[.25pt]\\[-1.85ex]
$\Gamma$&$1.70_{-0.04}^{+0.04}$&$2.15_{-0.20}^{+0.21}$&$1.29_{-0.04}^{+0.08}$&$2.07_{-0.54}^{+0.37}$&$0.59_{-0.06}^{+0.06}$&$0.80_{-0.36}^{+0.50}$ & $1.56_{-0.03}^{+0.06}$ & ${2.22}_{-0.09}^{+0.10}$  & $1.56_{-0.04}^{+0.10}$ & ${2.26}_{-0.15}^{+0.16}$\\[0.5ex]

$E_{\rm fold}$ [keV] & $49_{-45}^{+53}$ & ${72}_{-16}^{+67}$  & $27_{-1}^{+1}$ & ${47}_{-23}^{+39}$  & $17.2_{-0.6}^{+0.7}$ &  ${16}_{-2}^{+6}$ & $33_{-1}^{+2}$& ${50}_{-3}^{+2}$  & $38_{-2}^{+5}$ & ${69}_{-16}^{+31}$ \\[0.5ex]

$E_{\rm cyc}$ [keV] & $55_{-4}^{+1}$ & ${51.6}_{-0.9}^{+0.8}$ & $56.4_{-0.8}^{+0.9}$ & ${52.1}_{-0.5}^{+0.6}$  & $58_{-1}^{+1}$ &  ${46}_{-2}^{+8}$ & $54.9_{-1.0}^{+0.8}$& ${50.7}_{-0.6}^{+0.3}$  & $54.1_{-0.7}^{+0.9}$ &  ${49.9}_{-1.3}^{+0.9}$  \\[0.5ex]

$\sigma_{\rm cyc}$ [keV] & $4.3_{-0.6}^{+0.7}$ & ${3.9}_{-1.0}^{+0.8}$ & $5.7_{-0.5}^{+0.6}$ & ${4.6}_{-0.5}^{+0.6}$  & $6.8_{-0.7}^{+0.7}$ & ${16}_{-9}^{+9}$ & $4.8_{-0.7}^{+0.7}$& ${0.18}_{-0.09}^{+0.69}$  & $4.5_{-0.7}^{+0.6}$ & ${3.8}_{-1.1}^{+0.8}$  \\[0.5ex]

$\eta_{\rm cyc}$ & $10.5_{-1.6}^{+1.8}$ & ${68}_{-36}^{+181}$ & $21_{-6}^{+6}$ & ${20}_{-6}^{+16}$  & $11_{-1}^{+2}$ & ${24}_{-9}^{+15}$ & $11.0_{-2.6}^{+3.4}$ & \textit{unconstrained} & $9.6_{-1.4}^{+1.5}$ & ${90}_{-48}^{+263}$ \\[0.5ex]

$E_{\rm bump}$ [keV]& -- & ${36}_{-3}^{+2}$  & -- & ${25}_{-2}^{+3}$  & -- & ${35}_{-5}^{+1}$ & -- & ${34.3}_{-0.7}^{+0.7}$ & -- &${38}_{-2}^{+1}$ \\[0.5ex]

$\sigma_{\rm bump}$ [keV]& -- & $12_{-2}^{+2}$  & -- & ${14}_{-5}^{+2}$  & -- & ${7.5}_{-1.2}^{+3.8}$ & -- & ${6.9}_{-0.5}^{+0.5}$& -- &${10}_{-1}^{+1}$ \\[0.5ex]

norm$_{\rm bump}$\tablefootmark{a} & -- & ${0.006}_{-0.002}^{+0.002}$  & -- & ${0.014}_{-0.004}^{+0.010}$  & -- & ${0.05}_{-0.04}^{+0.06}$ & -- & ${0.0015}_{-0.0002}^{+0.0002}$& -- &${0.003}_{-0.001}^{+0.001}$ \\[0.5ex]

${\chi}^{2}_{\rm red}/d.o.f.$ & $1.8/136$ & $1.0/133$ & $2.4/136$ & $0.7/133$ & $1.5/136$ & $1.0/133$ & $1.9/136$ & $0.8/133$ & $1.9/136$ & $0.9/133$ \\[0.5ex]
\bottomrule[.25pt]
\end{tabular}
\end{center}
\tablefoot{
\tablefoottext{a}{XSPEC normalization of the bump component, in units of photon\,keV$^{-1}$\,cm$^{-2}$\, s$^{-1}$ at $1\,$keV.}
}
}
\end{table*}

%% file: Chapters/6_Discussion.tex
\subsection{Timing results}\label{sec:time}
We have explored the behaviour of the pulse period in \src during its January--February 2012 
outburst using a pulse-phase connection technique. The measured period is used to construct 
pulse profiles in two energy bands: $3-20\,$keV (JEM-X), and $18-80\,$keV (ISGRI), 
as shown in Fig.~\ref{fig:isgripp} and Fig.~\ref{fig:jmxpp}.
An apparent decrease of the pulse period over the outburst is clearly seen.
The measured period derivative is $\dot{P}$ = ($-2.04\pm0.01$)~10$^{-7}$
(Fig.~\ref{fig:phaseconn} and Table~\ref{table:timing}). 
This trend is most probably due to a combination of the Doppler effect caused by the 
orbital motion of the neutron star and an intrinsic spin-up caused by the transfer of angular 
momentum from the accretion disk to the neutron star. \\

Considering the JEM-X ($3-20\,$keV) and the ISGRI ($18-80\,$keV) energy bands,
the pulse profiles of \src are strongly energy dependent 
(except in the very end of the outburst, i.e., Rev\# 1138), 
similar to the majority of accreting pulsars.
Thereafter, these two bands only will be considered to compare our pulse profiles results.
The ISGRI pulse profiles presented in Fig.~\ref{fig:isgripp} show two sharp peaks and a hint of a 
third peak in between (around phase $0.5$) which is not present in all observations. 
In the JEM-X pulse profiles, however, all three peaks are clearly visible 
in all observations, with the exception of Rev\# 1138.
Comparing the pulse profiles in the two energy bands, it appears that the most prominent of the peak 
in the JEM-X pulse profiles (peaking around phase $0.5$ in the top panel of Fig.~\ref{fig:jmxpp}) tends
to disappear at higher energies. While there is no commonly accepted interpretation of the pulse 
profile dependence on energy and luminosity, we note that a similar behavior is observed in Her~X-1 
\citep{2008A&A...482..907K}, 4U~0115+63 and V\,0332+63 \citep{Tsygankov_etal07}.
A qualitative interpretation of the pulse profile energy-dependence 
(based on the results of \citealt{2006A&A...451..187M})
has been proposed by \citet{Tsygankov_etal07} 
to explain the variation of V\,0332+63 pulse profiles in a purely geometric fashion: 
if the axis of the magnetic dipole is at an angle to the rotation axis 
(a necessary condition to observe pulsation), 
the neutron star itself can eclipse some of the lower parts of the accretion columns, 
which are thought to primarily emit the harder photons.
Thus, compared to the pulse profile at softer energies, the harder one lacks of (at least) one peak.
On the contrary, the softer photons are primarily from regions higher up in the accretion column, 
and therefore visible during most of the spin period.
On the other hand, gravitational light deflection also plays 
a crucial role in shaping the observed pulse profile.
Indeed, depending on to the geometry of the emitting region and the observer's line of sight, 
this effect could make the accretion columns visible from all directions \citep{Kraus01, Kraus+03},
thus affecting the pulse profile in a significant way.
In addition, the pulse profiles luminosity dependence 
probably reflects a change in the physical condition of the accretion column. 
Indeed, for sub-critical sources, the height of the accretion column is anti-correlated with the accretion rate, 
and this could be at least partially responsible for the luminosity dependence of the pulse profiles. 
However, this is only a qualitative scenario to account for the observed pulse profile changes. 
More accurate descriptions of the pulse profiles dependences need more complicated models in order to account 
for the magnetic field configuration, the scattering cross section along the accretion column, 
the shape of the emission beam and its dependence on luminosity and energy, and other parameters. 
On the other hand, other accreting X-ray pulsars show the opposite behavior, where the low-energy 
pulse-profiles are absorbed, and the extra peak only appears at higher energies 
(see, i.e., \citealt{2004A&A...427..975K}, for the case of GX~301-2). 

%Fig. 10
\begin{figure}[t]
\includegraphics[width=\hsize]{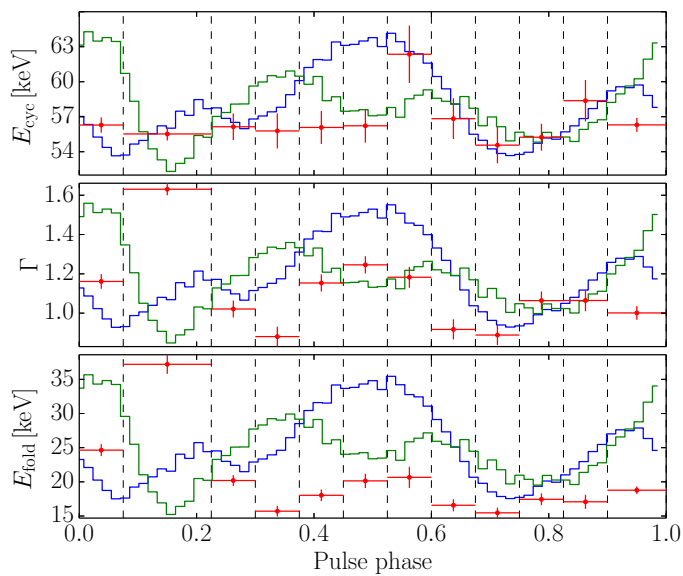}
\caption{The best fit parameters of the phase-resolved spectra using the five stacked INTEGRAL observations.
  The vertical dashed lines indicate the phase-bins. 
  The solid blue curve is the JEM-X average pulse profile, while the green curve is the ISGRI average pulse profile.
  The vertical error bars are $1\sigma$ uncertainties, while the horizontal bars represent the phase bin width.}
\label{fig:totprs}
\end{figure}

Although the overall morphology of the profile remains the same in all observations 
(which permits phase connection of the observations, see Sec.~\ref{sec:prs}), 
the relative intensities of the peaks slightly vary with luminosity.
Furthermore, in the last stage of the outburst (Rev\# 1138), the pulse profile resembles a single broad peak, 
and a weak energy dependence.
A similar flux and energy dependence of the pulse profile is also observed by \citet{2011MNRAS.417..348D} 
during a \src outburst in August $2010$, when the source reached a peak flux of $\sim700\,$mCrab 
(that is, ${\sim}30\%$ less than the outburst analyzed in this work).
However, the different pulse profiles of Rev\# $1138$ likely indicate a possible change 
in the beaming pattern from an almost purely pencil beam at lowest luminosities, 
to a mixed pencil-fan beam as the luminosity increases (see \citealt{Becker+12}, and references therein).

The pulsed fraction is only ${\sim}20\%$ in both energy bands, and shows only slight variation 
along the outburst, uncorrelated with the flux.
This is contrary to many other X-ray pulsars, where a pulsed fraction $>50\%$ is observed,
correlated with energy and flux \citep{Lutovinov+Tsygankov08}.
This again may be understood in a qualitative manner following a geometrical interpretation:
it is possible that, as the star rotates, we are exploring only a small area of the 
beam pattern (i.e., the observed flux from one emission region 
as a function of the direction to the observer), 
which in turn leads to an observed low pulsed fraction.
This scenario is supported by the results discussed in Section~\ref{sec:prs}.

\subsection{Phase-averaged spectral results}\label{sec:discspe}

With the new INTEGRAL calibration \citep{CaballeroISGRI+13} and analysis software (OSA~10), 
we confirm the luminosity-dependence of the cyclotron line energy $E_{\rm cyc}$ in \src reported 
in \citet{Klochkov+12}, with a better restriction of the spectral parameters.
The observed $E_{\rm cyc}$ shows a systematic shift of ${\sim}5\%$ with respect to the energies 
reported in the previous work.
As a consequence, the correlation found in this work has a slightly higher slope than that found by 
\citet{Klochkov+12}.
As mentioned in Section $1$, such dependency has already been found in several cyclotron line sources. 
The $E_{\rm cyc}$--flux correlation is positive in Her X-1 \citep{Staubert+07}, 
Vela X-1 \citep{Fuerst+14}, A\,0535+26 \citep{Sartore+15}, 
and some indications suggest also in GX~301-2 \citep{2005A&A...438..617L}. 
Pulsars with a confirmed negative $E_{\rm cyc}$--flux correlation 
are V\,0332+63 \citep{Tsygankov_etal06}, and 4U~0115+63 where the negative correlation was found 
using the NPEX model (\citealt{Nakajima_etal06}), 
but not when using a cutoff-power law model \citep{SMueller_etal13},
or when the fundamental is modelled by two independent sets of CRSF lines 
at ${\sim}11\,$keV and ${\sim}15\,$keV \citep{Iyer+15}. 
Thus, \src further enlarges the group of sources showing the positive correlation.

\citet{Klochkov+12} interpret this correlation according to the model described 
by \citet{Staubert+07} and \citet{Becker+12}.
In these models, a critical luminosity $L_c$ discriminates between two accretion 
regimes: one in which the height of the emitting column is governed by the ram pressure 
of the infalling material (for $L<L_c$, i.e., ``sub-critical sources''), 
and the other where a radiation pressure dominated emitting column is formed
(for $L>L_c$, i.e., ``super-critical sources'').
As the luminosity $L$ varies, the X-ray emitting height also moves 
within the accretion column, thus the local magnetic field changes, 
leading to a variable cyclotron line energy with luminosity.
The critical luminosity below which a positive $E_{\rm cyc}$--luminosity correlation is expected,
has been calculated by \citet{Becker+12} as a function of the magnetic field 
and other physical parameters of the NS.
For typical neutron star parameters, it results in $L_c\sim 1.5\times10^{37} B_{12}^{16/15}\,$erg~s$^{-1}$. 
For GX~304-1, this leads to $\sim8\times10^{37}\,$erg~s$^{-1}$.
During the analyzed outburst, the source luminosity clearly remains below this value 
(see Figs. \ref{fig:OSA10} and \ref{fig:gamma}). 
Thus, the observed positive $E_{\rm cyc}$--flux correlation is consistent with the expected behavior 
for a sub-critical source according to the physical picture outlined in \citet{Becker+12}.
More recently, a new calculation of the $L_c$ function has been proposed by \citet{Mushtukov+15},
where a more detailed accretion scenario is taken into account.
The observed positive $E_{\rm cyc}$--flux relation in \src is consistent with their model as well.

\begin{figure}[!t]
\label{fig:HD}
\includegraphics[width=\hsize]{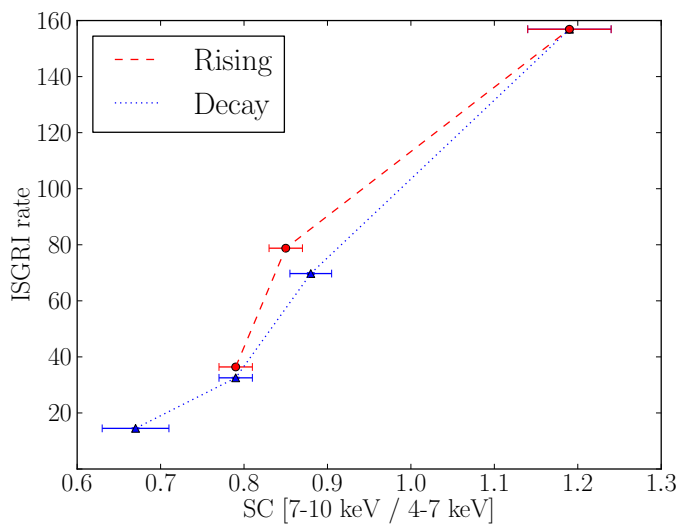}
\caption{Hardness/intensity diagram for \src during the analyzed outburst. 
The soft color definition is given in the text. 
The errors are 1$\sigma$ uncertainties.
The errors on the ISGRI rates are smaller than the symbol size. 
The dashed red curve corresponds to the rising part of the outburst, the dotted blue one -- to the decay.}
\label{fig:HD}
\end{figure}

We have also investigated the correlation between the spectral photon index $\Gamma$ and luminosity.
In many accreting X-ray pulsars, $\Gamma$ is luminosity dependent. 
It had been found that in accreting pulsars showing a positive $E_{\rm cyc}$--flux correlation, 
the correlation between $\Gamma$ and the flux is negative and vice versa \citep{2011A&A...532A.126K}. 
The exact mechanism of the continuum/flux dependence is not yet clear, 
especially for super-critical pulsars (with $L>L_{\rm c}$). 
In sub-critical sources, including \src, the observed hardening of the spectrum with flux 
(Fig.~\ref{fig:gamma}) can be qualitatively understood assuming that the pressure and electron density 
inside the emitting structure increase with increasing mass accretion rate and luminosity. 
Indeed, such an increase should lead to a higher optical depth $\tau$, thus to a 
larger Comptonization parameter $y=\tau kT_e/(m_ec^2)$.
The spectrum formed by Comptonization of the seed photons from the base of the column would then become harder.

Finally, we have investigated the dependence of the folding energy 
$E_{\rm fold}$ on the photon index $\Gamma$ and, at the same time, on luminosity.
These correlations can be seen in the contour plot in Fig.~\ref{fig:contours}.
The plot shows that $E_{\rm fold}$ is positively correlated with $\Gamma$,
and negatively correlated with flux ($E_{\rm fold}$ increases at
lower luminosities). 
A similar behavior is reported for other two accreting pulsars:
A~0535+26 \citep{MuellerD+13} and RX~J$0440.9$+$4431$ \citep{Ferrigno+13}.
These correlations can be qualitatively understood as follows.
It is generally accepted that the powerlaw--cutoff spectrum of accreting pulsars 
is a result of Comptonization, i.e., scattering of soft photons off hot electrons
supplied by the infalling plasma.
As the luminosity increases towards $L_{\rm c}$, radiation field begins to affect 
the accretion flow and Compton cooling becomes more efficient.
This might lead to the observed shift of $E_{\rm fold}$ reflecting the electron 
temperature $T_e$ towards lower energies while, at the same time, 
the Comptonization process makes the spectrum harder.

The dependence of the spectral hardness on flux has recently been investigated 
for a number of accreting pulsars in BeXRBs by \citet{2013A&A...551A...1R}. 
They found that the X-ray emission of the sources follows certain patterns in the hardness/intensity diagram. 
At lower fluxes, the emission first becomes harder with increasing flux. 
At a certain flux a ``turn-over'' occurs, so that at higher fluxes the emission 
again becomes softer (Fig.~\ref{fig:gamma} in \citealt{2013A&A...551A...1R}). 
The luminosities corresponding to the fluxes of the turn-over are in all cases close 
to the critical luminosity $L_{\rm c}$ discussed above. 
Thus, the turn-overs in the hardness/flux dependences may reflect a transition between the 
aforementioned sub- and super-critical regimes characterized by the opposite $E_{\rm cyc}$--flux correlations. 
To compare our results with those presented in \citet{2013A&A...551A...1R}, we constructed 
a hardness/intensity diagram for \src (Fig.~\ref{fig:HD}). We use the same definitions of the 
``soft-color'' (SC) -- the ratio between the fluxes in the $7-10\,$keV and $4-7\,$keV ranges. 
As expected in a sub-critical source, the luminosity corresponding to the turn-over is not reached in \src 
and the diagram shows only the part where the hardness increases with luminosity, i.e., 
the ``horizontal branch'' according to the terminology used in \citet{2013A&A...551A...1R}.

\subsection{Phase-resolved spectral analysis}\label{sec:prs}

We modelled the phase-resolved spectra with a rolloff power-law model 
and a multiplicative Gaussian absorption line to account for the CRSF.
The maximum value of the line energy, $E_{\rm cyc}\sim62\,$keV, is reached at the peak of the outburst, 
Rev\#~$1134$, in the pulse phase bin 0.4--0.6 (see Table~\ref{table:prs}).
Using a value of $z_{g}=0.3$ for the gravitational redshift (assuming canonical neutron star parameters), 
this corresponds to a magnetic field strength of $6.9\times10^{12}\,$Gauss. 
This is among the strongest observed magnetic fields in an accreting X-ray pulsar. 

The centroid energy of the cyclotron line shows only slight variations with pulse phase, 
not exceeding ${\sim}16$\% (with respect to $0.2$ wide phase bins). 
This is contrary to some other X-ray pulsars (see, e.g., the cases of Cen X-3, \citealt{2000ApJ...530..429B}; 
GX~301-2, \citealt{2004A&A...427..975K}; and Her~X-1, \citealt{Klochkov+08, 2011A&A...532A..99V}), 
where the energy of the cyclotron line shows up to 30\% variation with pulse phase. 
\src, however, is not the only accreting X-ray pulsar where the cyclotron line energy 
varies only by $<20\%$ with pulse phase: see, e.g., the case of 4U~1538-52 \citep{Coburn_etal02}.

We find that the standard rolloff/power law model does not lead to good spectral fits for the 0.0--0.2 phase bins. 
These spectra need an additional component to obtain an acceptable fit.
We modelled this component with a Gaussian emission line around $35\,$keV.
Further analysis with more physically motivated models are needed to investigate the feature in more detail, 
and to understand its origin.
The extra component is reminiscent of the "$10\,$keV feature" observed in the spectra of many 
accreting X-ray pulsars \citep{Coburn_etal02, Klochkov+08, SMueller_etal13}, for which no accepted explanation of its 
physical origin exists. 
However, the centroid energy of the feature in \src is at a much higher energy, $\sim35\,$keV.
Interpreted within this context, the feature is likely due to the inadequacy of the underlying phenomenological model.
Alternatively, we found that the same feature can be fitted by a Gaussian absorption line at
lower energies, instead of an emission line.
In this case, the absorption line centroid energy is between $12\,$keV (with $\sigma=14\,$keV, $\eta=37$),
and $25\,$keV (with $\sigma=9\,$keV, $\eta=8$).
To explore the possibility of the feature being the fundamental cyclotron line,
we calculated the harmonic ratio between the confirmed cylotron line at ${\sim}55\,$keV, 
assumed to be the first harmonic, and the centroid energy of the absorption feature, 
and found that it significantly deviates from the canonical ratio of 2.
For the feature at $12\,$keV, the ratio results in an anomalously high value of ${\sim}4.6$, 
while for the feature at $25\,$keV the ratio is halved to a value of ${\sim}2.3$.
Relatively high deviations from an integer value of the harmonic ratios are observed also in other sources, 
e.g., Vela~X-1 (\citealt{Fuerst+14}, and references therein), 
where the ratio between the confirmed harmonic and fundamental line energies spans in the range $2.1$--$2.7$.
Those authors invoke photons spawning from the harmonic lines 
as responsible to influence the shape of the fundamental line,
shifting its measured centroid energy from its real value.
However, values of the harmonic ratio as high as $4.6$ have never been observed 
for the first harmonic line, and seem to exclude the possibility 
of the absorption feature being the fundamental cyclotron line.

Thanks to the successful phase-connection (see Section~3), 
we have been able to identify phase bins at different stages of the outburst and, therefore, 
to perform phase-resolved spectroscopy in order to explore the spectral variability with pulse phase.
In particular, we have been able to perform perform phase-resolved spectroscopy 
stacking together all observations (with the exception of Rev\# 1138), 
in order to improve the statistics and investigate the pulse phase 
variability of spectral parameters with better precision ($12$ phase bins).
The results are shown in Fig.~\ref{fig:totprs} for three parameters of interest:
the cyclotron line centroid energy, the photon index and the folding energy.
To compare the phase-resolved parameters of the stacked analysis, with the pulse profiles,
we produced averaged JEM-X and ISGRI pulse profiles, which are also shown in Fig.~\ref{fig:totprs}.
The value of $E_{\rm cyc}$ is constant at $\sim56\,$keV 
(with one phase bin showing a ${\sim}10\%$ higher value, however, not statistically significant).
The photon index $\Gamma$ and the folding energy $E_{\rm fold}$ show more prominent 
variations with pulse phase, however, less than a factor of ${\sim}2$.
A similar variation of the continuum parameters is observed also in other accreting pulsars
(see, e.g., Her~X-1, \citealt{Vasco+13}; GX~301-2, \citealt{Suchy+12}; Vela~X-1, \citealt{MaitraVela13}),
but a consistent theoretical scenario which supports such variations is still missing.

Similarly to the phase-resolved analysis of the separated observations, 
a few phase-bins of the stacked phase-resolved spectra need 
an additional component to give an acceptable $\chi^2$.
We modelled this component by a Gaussian emission line around $35\,$keV.
An equally good fit is obtained with an absorption Gaussian line at lower energies, i.e., 
between $16\,$keV (with $\sigma=9\,$keV, $\eta=17$),
and $21\,$keV (with $\sigma=13\,$keV, $\eta=24$).

Since our timing analysis shows a low pulsed fraction which is flux- and energy-independent, 
and the stacked phase-resolved analysis shows a remarkably stable $E_{\rm cyc}$, 
we are tempted to explain these results within the context of the same scenario.
Indeed, the steadyness of the cyclotron line energy along the pulse phase can be understood as follows.
According to model calculations, the cyclotron line parameters should strongly depend 
on the viewing angle and, therefore, on the pulse phase (see \citealt{Schoenner+07}, and references therein).
However, if the geometrical configuration is such that we are exploring only a small portion 
of the beam pattern, then the parameters result mostly insensitive to the pulsar rotation.
Therefore, the viewing angle with which we observe the accretion column does not 
vary appreciably with the NS rotation and, as a consequence, 
the $E_{\rm cyc}$ results almost independent from the rotational phase.

%% file: Chapters/7_Conclusions.tex
In this work, we present the results from spectral and timing analysis of INTEGRAL data of the 
accreting pulsar \src observed during an outburst in January-February $2012$.

We studied the behavior of the source spectrum as a function of the luminosity analyzing the 
pulse phase-averaged and phase-resolved spectral parameters. Our results confirm the positive 
$E_{\rm cyc}$-flux correlation in \src with the new INTEGRAL calibration, and show a negative 
correlation between the photon index and the luminosity. 
The positive $E_{\rm cyc}$--flux correlation is also observed in other accreting pulsars 
whose luminosity is below $L_c$, like Her~X-1 and Vela~X-1, while the negative photon index-luminosity 
correlation has also been observed in the pulse-to-pulse analysis of Her~X-1 and A~0535+26. 
Then, \src is one of the few sub-critical sources with a confirmed photon index-luminosity 
negative correlation. 
To test whether the behavior of \src below $L_c$ in a model-independent way, we constructed a 
hardness/intensity diagram and verified that the turn-over point, which reflects the
critical luminosity, is never reached, and the diagram only shows the so-called horizontal branch.
Furthermore, a correlation between the folding energy and the photon index has been found,
with a relative anti-correlation between the folding energy and the observed luminosity.
These correlations seem to fit well in the context of an emergent spectrum shaped by 
both bulk and thermal Comptonization of blackbody seed photons radiating from the accretion mound
(where thermal Comptonization becomes important when the luminosity increases towards $L_{\rm c}$).
In this context, the observed spectral changes are a consequence of the increasing 
Compton cooling efficiency at higher luminosity.

Thanks to the pulse profiles main features, and to their clear identification in all observations (except the last one), 
we have been able to phase-connect the pulse profiles, which allowed us to 
obtain a pulse period solution valid over the entire outburst with remarkable precision, 
and to identify phase bins at different stages of the outburst.
We measured a period derivative seen over the full outburst, which is probably a combination 
of the Doppler shift due to the orbital motion of the neutron star and an intrinsic spin-up 
due to the angular momentum transfer by the accreted material.

The observed pulse profiles compared in the JEM-X ($3-20\,$keV) and ISGRI ($18-80\,$keV) energy bands 
show a strong energy and luminosity dependence during the outburst, except at the end of the outburst (Rev\# 1138), 
where the pulse profiles are similar in the two energy bands and show a much less complex morphology.
While the energy dependent pulse profiles can be justified by a geometrical effect due to the 
rotation of the neutron star, the luminosity dependence probably reflects changes in 
the physical details of the accretion process in the sub-critical regime.
Moreover, we observe only a low pulsed fraction, independent of energy and luminosity.

The phase resolved analysis shows that the cyclotron line is detectable at all rotational phase 
bins of the neutron star. 
The maximum reached centroid energy (${\sim}62\,$keV) enabled us to infer one of the 
strongest magnetic field strength on an accreting X-ray pulsar 
(while the highest belongs to GRO~J$1008$-$57$ with a cyclotron line energy of ${\sim}80\,$keV, \citealt{Bellm+14}).

Furthermore, we observe no significant variations of the centroid energy with the pulse phase.
On the other hand, the depth and the width of the line, show more pronounced variations 
with pulse phase, up to a factor of ${\sim}3$, with no correlation with the pulse profile.
In the stacked phase-resolved analysis we have been able to show how the cyclotron line energy is 
essentially constant along the pulse phase.

All these considerations lead us to a possible qualitative scenario where the geometrical configuration
of the observed NS is such that we are able to explore only a small part of the beam pattern.

We found that, contrary to the phase averaged spectra, a few phase-resolved spectra lead 
to large residuals when fitted by a standard rolloff/power-law model: they need an additive 
component that we modeled by a Gaussian emission line around ${\sim}35\,$keV, 
but that can be equally modeled by an absorption line at ${\sim}20\,$keV.
The additional Gaussian emission line heavily modifies the resulting best-fit parameters, 
often leading to unphysical values, most likely indicating that the feature is due to the 
inadequacy of the underlying phenomenological model.

%% file: Chapters/Appendix.tex
\label{sec:appendix}

\begin{figure}[!h]
    \includegraphics[width=\hsize]{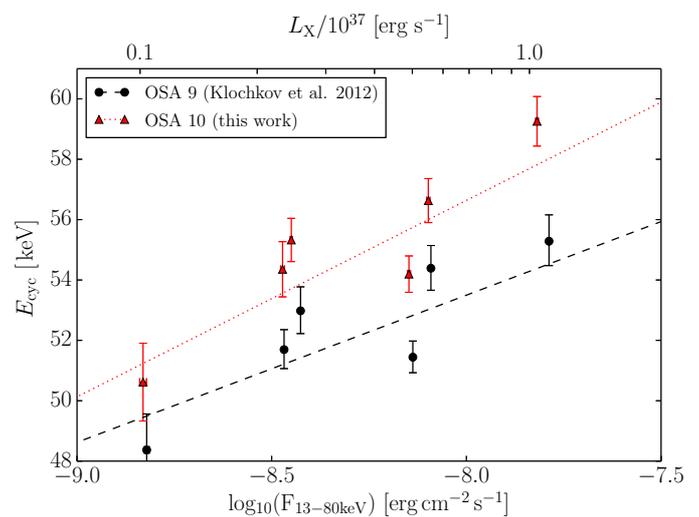}
    \caption{Cyclotron line centroid energy $E_\mathrm{cyc}$ as a function of the logarithm of flux in the $13-80\,$keV range. 
    The black dots are the results of \citet{Klochkov+12}, obtained with OSA~9 and corrected for the ISGRI energy gain drift. 
    The red triangles are the results of this work, obtained with OSA~10 and further gain corrected.
    The dashed and dotted lines are the results of linear fits to the $E_\mathrm{cyc}-log_{10}(Flux)$ data 
    for the two sets resulting from OSA~9 and OSA~10 data reduction, respectively.
    The error bars indicate $1\sigma$-uncertainties (the statistical flux uncertainties are smaller than the symbol size).
    The top x-axis shows the corresponding isotropic source luminosity assuming a distance of $2.4\,$kpc.}
\label{fig:OSAs}
\end{figure}

Our spectral analysis has been performed with the OSA~10 software which,
at the time of writing, is the latest version of the INTEGRAL analysis software.
OSA~10 is supplied with a new energy calibration of the ISGRI detector, with respect to the previous version (OSA~9).
As a result, a more reliable photon energy reconstruction is ensured.
The details about the new energy reconstruction are reported in \citet{CaballeroISGRI+13}.

To investigate the difference in energy calibration between OSA~9 and OSA~10, 
we extracted the data set used in this work with both the software versions.
Following \citet{Klochkov+12}, after the OSA~9 data reduction, we performed an additional gain correction
based on the background spectral line complex of Tungsten (W, nominally at $58.8297\,$keV).
After the OSA~10 data reduction, which includes a more accurate gain correction,
a small deviation of the inferred energy from the nominal Tungsten line was still noticeable.
The deviation amounted up to $+0.5\,$keV in the Science Windows (ScWs, i.e., INTEGRAL pointings)
at the beginning of each INTEGRAL revolution.
According to the ISGRI instrument team (priv. comm.), the deviation can be attributed to 
possible orbital drifts in the detector response, which are responsible for systematic scatter when revolutions start.
However, even if orbital variations are not currently corrected, they are mitigated when averaged over a number of revolutions.
Nonetheless, to eliminate this deviation, we performed an additional gain correction, 
in order to ensure the stability of the Tungsten reference line.
After the correction, the inferred Tungsten line energy in all ScWs is found to be 
consistent with its nominal value within $3\sigma$.

To illustrate the effects of the different energy calibrations, 
we plotted the $E_{cyc}$-flux relation for the spectra obtained with the two software versions (Fig.~\ref{fig:OSAs}).
We note that the energy range indicated in \citet{Klochkov+12} for the calculation of the flux 
is erroneously reported as $4-80\,$keV, while the correct range is $13-80\,$keV 
(where the lower limit is constrained by the response matrix energy range of the ISGRI instrument).
The two different sets of results clearly show a systematic shift of the cyclotron line centroid energy,
with a deviation (proportional to the measured energy) up to ${\sim}2\,$keV.
This is the result of the new energy calibration implemented in OSA~10 plus a minor contribution 
due to the additional gain correction performed after the OSA~10 data reduction.
In our scientific analysis, we used the data reduced with OSA~10 with the additional gain correction.
This ensures the most accurate measurements of the cyclotron line centroid energy.